\documentclass{llncs}
\usepackage{multirow}
\usepackage{multirow}
\usepackage{graphicx}
\usepackage{subfigure}
\usepackage{amsfonts}
\usepackage{amsmath}
\usepackage{algorithmic}
\usepackage{algorithm}

\begin{document}

\title{Probabilistic Latent Tensor Factorization \\
Model for Link Pattern Prediction \\
in Multi-relational Networks}

\author{Sheng Gao, Ludovic Denoyer and Patrick Gallinari}

\institute {LIP6 - UPMC \\
            4 place Jussieu, 75005 Paris, France}

\maketitle

\begin{abstract}
This paper aims at the problem of link pattern prediction in collections of objects connected by multiple relation types, where each type may play a distinct role. While common link analysis models are limited to single-type link prediction, we attempt here to capture the correlations among different relation types and reveal the impact of various relation types on performance quality. For that, we define the overall relations between object pairs as a \textit{link pattern} which consists in interaction pattern and connection structure in the network, and then use tensor formalization to jointly model and predict the link patterns, which we refer to as \textit{Link Pattern Prediction} (LPP) problem. To address the issue, we propose a Probabilistic Latent Tensor Factorization (PLTF) model by introducing another latent factor for multiple relation types and furnish the Hierarchical Bayesian treatment of the proposed probabilistic model to avoid overfitting for solving the LPP problem. To learn the proposed model we develop an efficient Markov Chain Monte Carlo sampling method. Extensive experiments are conducted on several real world datasets and demonstrate significant improvements over several existing state-of-the-art methods.
\end{abstract}

\section{Introduction}
Modeling relational data has been an active area of research in recent years, and is becoming an increasingly important problem in many applications such as social network analysis and recommender systems. Link prediction \cite{getoor:survey} as one basic challenge is concerned with predicting unobserved links between object pairs based on the observed structure in the network. A typical example is a social network where people are linked via explicit relations, such as friendship or membership; or implicit ones like the sharing of similar interests. Up to now, most of the related models developed for link prediction either consider only single-type relations among objects or treat the different relations in the network homogeneously \cite{Acar:Link} \cite{Liben-Nowell:The}, thus ignoring the multi-dimensional nature of interactions and the potential complexity of the interaction schemes in the networks.

In this paper, we focus on the task of predicting multiple relation types among object pairs in multi-relational networks. For that, we define the overall relations between each pair of objects as a \emph{link pattern}, which consists in interaction pattern and connection structure among objects. This task is illustrated in Figure 1. The left part of Figure 1 shows a social network composed of set of individuals with multiple relations among them, where the link patterns involving multiple relations between certain object pairs are unobserved (indicated by $"?"$ in the figure). The task here is to infer the missing link patterns from the observed part of the network, which we refer to as \emph{Link Pattern Prediction} (LPP) problem. With the extracted link patterns, the fine and subtle network structure in the multi-relational networks can be captured effectively, and can be used to improve the range and performance of various aspects of social network applications, including community detection and person recommendation with different social roles.
\begin{figure}[t]
\centering
\includegraphics[width=4in]{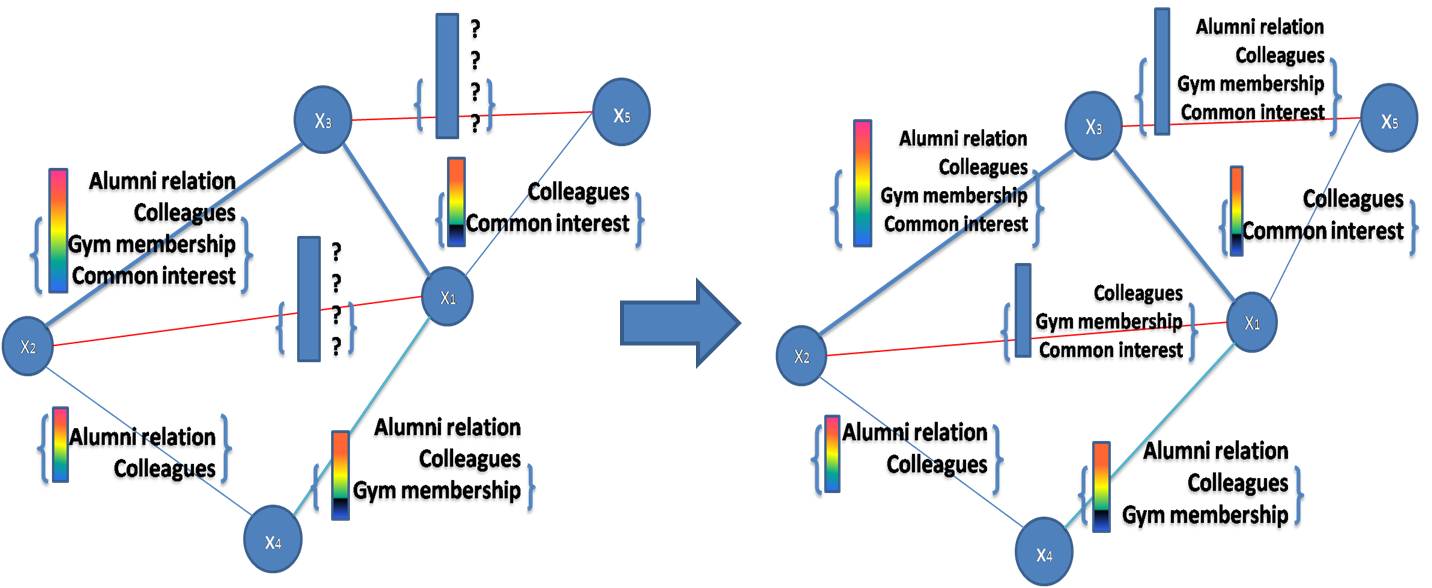}
\caption{Example of link pattern prediction task. There are four relation types (Alumni relations, Colleagues, Gym membership and Common interest in the social network. The sets of $"?"$ represent unknown link patterns.}
\end{figure}

Therefore, in the context of Link Pattern Prediction problem, we propose a probabilistic tensor factorization framework to model the multi-relational data by considering the tensor factorization as a latent factor model. Our model addresses two major challenges that are ignored by previous work on link prediction. The first challenge is the multi-relational nature of networked data. In addition to using latent factors to characterize object features, we also introduce another latent factor for different relations to capture the correlations among multiple relation types and reveal the impact of distinct relation types on performance quality. The second challenge is data sparsity problem. For example, the social networks are usually very sparse, and the presence of relations among users only hold a very small number of all possible pairs of nodes. To solve the overfitting problem caused by the sparse data we extend our probabilistic model by employing Bayesian learning method to infer the latent factors. The Bayesian treatment can significantly avoid overfitting by placing prior information on the model parameters and handling the missing data easily.

Moreover, we deal with the parameter learning by an efficient Markov Chain Monte Carlo (MCMC) method in the real world datasets. We conduct experiments on several real world multi-relational datasets, the empirical results demonstrate the improvement of prediction accuracy and the effectiveness of our models.

The rest of the paper is structured as follows. We first briefly review related work in Section 2. Then we introduce our link pattern prediction task and formulate the probabilistic latent tensor factorization model for solving the LPP problem in Section 3. We also provide a fully Hierarchical Bayesian treatment to optimize the probabilistic model and derive an efficient Markov Chain Monte Carlo optimization method in Section 4. We describe experiments on three real-world datasets to study the efficiency of our model and compare it to several models in Section 5. In Section 6 we present conclusions and future work.

\section{Related Work}
Previous link prediction models are entirely based on structural properties of the observed network. \cite{Liben-Nowell:The} compares many predictors based on different graph proximity measures. Then the latent factor models have received more and more attention in recent years. \cite{y:dlp} propose the stochastic relational models for the link prediction problem, which are essentially the Gaussian process models. \cite{r:pmf} extend the matrix factorization model to the probabilistic framework for collaborative filtering tasks. Another interesting direction on link prediction concerns the prediction of the relationship strength \cite{Kahanda:Using}. However, all the related work above in link prediction literature focus only on the single link prediction task.

The problem of link prediction in multi-relational networks has only been addressed very recently, e.g. \cite{Kashima:Link} \cite{c:clp}. Our work is related to the multi-relational learning problem, where several relations are jointly modeled. Different strategies have been developed to enable parameter sharing when jointly factorize a collection of related matrices. For example, \cite{kurt:lfrm} introduced the nonparametric latent feature relational model which infers only the latent features of each entity. \cite{hoff:mlf} proposed a statistical model with latent matrix representations of the objects and the regression term to predict the missing links in the network data without considering relation types. However, our proposed model additionally introduce the latent factor for multiple relation types to capture the correlations and explore the impact of distinct relation types on prediction performance.

Tensor factorization has also attracted a lot of attention in the data mining community and have been used in many applications, such as web link analysis \cite{Kolda:Higher} \cite{Acar:Link}, analysis of email communications \cite{b:discussion}, and for personalized tag recommendations \cite{steffen:tag} or collaborative filtering over time \cite{x:tcf}. There are also other tensor factorization models \cite{p:hdp} \cite{m:nft} which are not suitable for our proposed LPP problem.

\begin{figure}[t]
\centering
\includegraphics[width=4in]{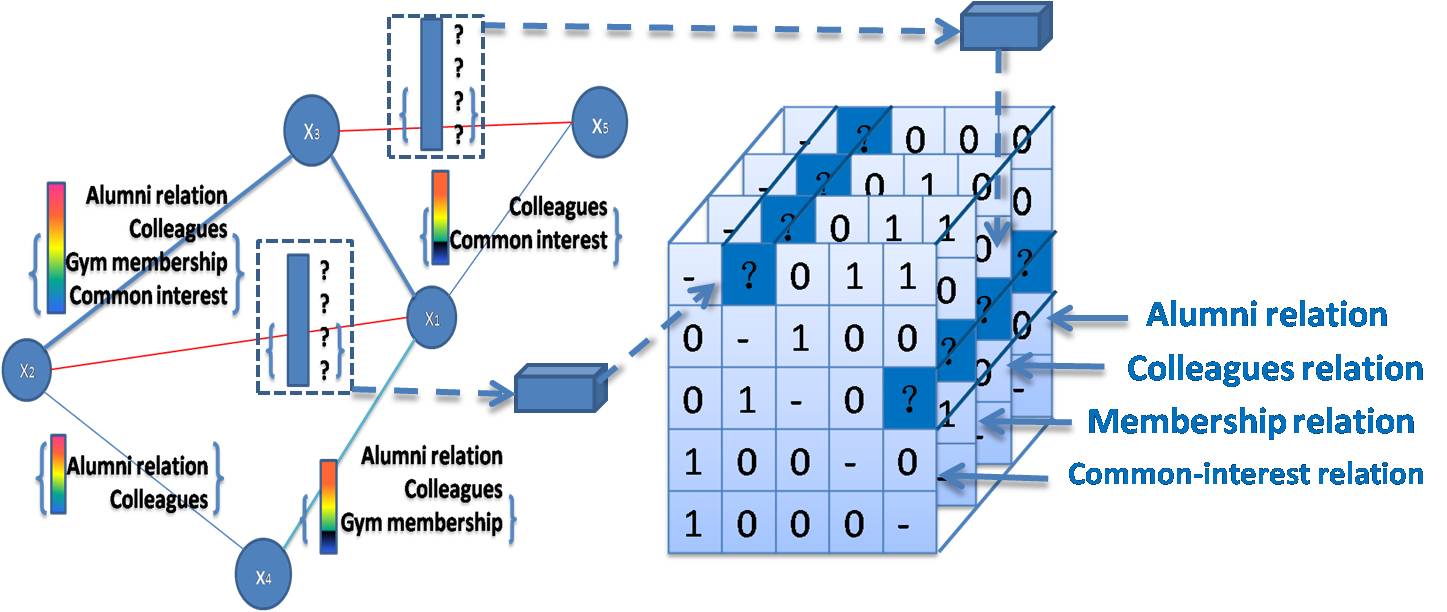}
\caption{Example of modeling the multi-relational social network as a tensor for predicting the unobserved link patterns in the network. To the right is the tensor representation of network where each slice matrix represents one relation type and each tube fiber represents the link pattern between two nodes. Unknown link patterns are represented as $"?"$ fibers.}
\end{figure}

\section{Probabilistic Latent Tensor Factorization Model for Link Pattern Prediction}
In this section, we first present problem definition for the link pattern prediction task, then propose a latent tensor factorization framework to model the multi-relational data and develop a \emph{maximum a posteriori} (MAP) method to infer the latent factors.

\subsection{Problem Definition}
Suppose we are modeling pair-wise relations between objects. Formally, let $X={\{{x_{1},{x_{2}},{\cdots}{x_{N}}\}}}$ represents a set of \textit{N} objects, assuming there are \textit{T} different types of relations among object pairs defined by a set of \textit{T} adjacency matrices. Based on the multi-dimensional nature of tensor representation, we model the multi-relational data in the network by an ${\textit{N}}{\times}{\textit{N}}{\times}{\textit{T}}$ third-order tensor $\mathcal{Y}$, where an entry $\mathcal{Y}_{ijt}$ representing the t-type relation value between object pair ($x_i$,$x_j$) can be defined as follows:
$ \forall i,j \in [1..N]^2, \forall t \in [1..T],$
\begin{equation*}
\mathcal{Y}_{ijt} = \left\{
\begin{array}{ll}
      1  & \hbox{if a t-type relation exists for $x_{i}$ and $x_{j}$;} \\
      0  & \hbox{if no t-type relation exists for $x_{i}$ and $x_{j}$;} \\
      ?  & \hbox{if the relation for $x_{i}$ and $x_{j}$ is unknown.}
\end{array}
\right.
\end{equation*}

We also define another ${\textit{N}}{\times}{\textit{N}}{\times}{\textit{T}}$ third-order tensor $\mathcal{I}$ which serves as the indicator tensor that is equal to 1 if the link is observed and equal to 0 when the link status is missing for the pair of objects.

Then, we define $\mathcal{Y}_{(ij:)}$ as the \emph{link pattern} involving \textit{T} different types of relations between each pair of objects $x_{i}$ and $x_{j}$, which is naturally represented by a tube fiber in the tensor model. Given some observed link patterns information for the related objects in the multi-relational network, we are interested in the task of predicting the unobserved link patterns for the remaining object pairs as illustrated in Figure 2, which we refer to as Link Pattern Prediction (LPP) problem.

\subsection{Probabilistic Latent Tensor Factorization Model}
It has been shown that missing links can be inferred by the inner product of corresponding latent features based on matrix factorization \cite{r:pmf} \cite{kurt:lfrm}. However, in LPP problem the link patterns generate multi-dimensional data which are difficult to handle by matrix factorization. Motivated by this observation, we extend the Probabilistic Matrix Factorization \cite{r:pmf} to the tensor factorization version and propose a Probabilistic Latent Tensor Factorization (PLTF) model to explore the latent factor matrices and discover the unobserved link pattern information.

Assuming the proposed PLTF model assigns different latent factor matrices for the pair of related objects, denoted as $U \in \mathbb{R}^{N \times D}$, $V \in \mathbb{R}^{N \times D}$, and a specific latent feature factor for the relation types, denoted as $R \in \mathbb{R}^{T \times D}$, then the probability of having a relation between objects can be learned via the three-order tensor factorization on data $\mathcal{Y}$.

In the work we consider CANDECOMP/PARAFAC(CP) tensor factorization \cite{s:lpp} \cite{j:analy} for the data modeling, and introduce the generalized tensor factorization model which can be written as follows:
\begin{equation}
\mathcal{Y}_{i,j,t} \sim f(\sum\limits_{d=1}^D U_{i,d} V_{j,d} R_{t,d} + E)
\end{equation}
where $E$ is the tensor form of noise term.

By assuming the noise term $E$ be Gaussian, the observed data $\mathcal{Y}$ \footnote {For convenience, we just use $\mathcal{Y}$ representing the observed part of data $\mathcal{Y}_{ijt}$} in the generalized tensor factorization model would follow a multivariate Gaussian distribution as follows:
\begin{equation}
\begin{aligned}
p({\mathcal{Y}}| U, V, R,\alpha) = \prod\limits_{i=1}^N \prod\limits_{j=1}^M \prod\limits_{t=1}^T \mathcal{I}_{ijt} \big[ \mathcal{N}({\mathcal{Y}_{ijt}} |(\sum\limits_{d=1}^D U_{i,d}V_{j,d}R_{t,d}), \alpha ^{-1}) \big]
\end{aligned}
\end{equation}
where $\mathcal{N}(\cdot)$ denotes the multivariate Gaussian distribution with precision $\alpha$ and mean $ ( \sum_{d=1}^D U_{i,d}V_{j,d}R_{t,d} )$ obtained by the CP tensor factorization.

Following the process on predicted data in \cite{r:pmf}, we also consider to incorporate the logistic function $g(x)= \frac{exp(x)}{1+exp(x)}$, which can be used to map the product of latent factors $U$, $V$ and $R$ for missing link pattern prediction results into the range [0,1]. Then we can obtain the transformed predictive distribution rewritten as follows:
\begin{equation}
\begin{aligned}
\label{repredpltf}
p(\mathcal{Y}| U, V, R,\alpha) = \prod\limits_{i=1}^N \prod\limits_{j=1}^M \prod\limits_{t=1}^T \mathcal{I}_{ijt} \big[ \mathcal{N} \big( {\mathcal{Y}_{ijt}} |g(\sum\limits_{d=1}^D U_{id}V_{jd}R_{td}), \alpha ^{-1} \big) \big]
\end{aligned}
\end{equation}
Note that by the transformation with the sigmoid function, we can somewhat improve the prediction performance.

\subsection{Basic Model with Non-informative Priors}
Considering the observed data in the tensor model is very sparse, we take the usual Bayesian model by imposing zero-mean and independent multivariate Gaussian prior distributions on the latent feature factor matrices $U,V$ for object pairs as follows:
\begin{equation}
\begin{aligned}
p(U \mid \alpha _{U}) = \prod\limits_{i=1}^N \mathcal{N}(U_{i} \mid 0, \alpha ^{-1}_{U} \mathbf{I})
\end{aligned}
\end{equation}

\begin{equation}
\begin{aligned}
p(V \mid \alpha _{V}) = \prod\limits_{j=1}^M \mathcal{N}(V_{j} \mid 0, \alpha ^{-1}_{V} \mathbf{I})
\end{aligned}
\end{equation}
where $\mathbf{I}$ is the D-by-D identity matrix.

In terms of the latent feature factor for relation types, we believe that in the contexts of different relation types the pair of objects will not have the same interaction pattern, which means they have independent feature vectors in the latent factor $R$. Therefore, we further place the similar Gaussian prior distribution on the latent factor $R$ for relation types as follows:
\begin{equation}
\begin{aligned}
p(R \mid \alpha_{R}) = \prod\limits_{t=1}^T \mathcal{N}(R_{t} \mid 0, \alpha ^{-1}_{R} \mathbf{I})
\end{aligned}
\end{equation}

Based on the prior distributions placed on the latent factors in the model, we can get the predictive distribution for the unobserved link patterns ${\mathcal{Y}_{(ij:)}^{*}}$ conditioned on observed data ${\mathcal{Y}}$ as:
\begin{equation}
\begin{aligned}
p({\mathcal{Y}^{*}}|{\mathcal{Y}}) = \int p(\mathcal{Y}^{*}|U,V,R,\alpha) p(U,V,R,\alpha | \mathcal{Y}) d\{U,V,R,\alpha,\alpha_{U},\alpha_{V},\alpha_{R}\}
\end{aligned}
\end{equation}

Considering the above model specification for the directed relations, we have an interpretation for modeling the probability of existing link pattern between object pairs: the link patterns involving multiple relation types among object pairs depend not only on how similar the sender-specific and receiver-specific latent feature factors $U$ and $V$ are, but also on how likely the corresponding latent features of object pairs match with the "context" latent factor $R$ representing different relation types. For example, in the same company one who likes sports may connect to the colleagues who are fellow gym-goers, and be less likely to interact with the alumnus that love shopping, which means the characteristics of people influence the probability of the link pattern in a number of different spheres of interaction or relation types.

\subsection{Parameter Estimation}
Considering the distribution in Equation (7), note that the predictive distribution is averaged over the posterior distribution $p(U,V,R ,\alpha| \mathcal{Y})$. We can infer the model parameters $\{U,V,R,\alpha \}$ by maximizing the log-posterior distribution over them as following:
\begin{equation}
\begin{aligned}
\ln p(U,V,R,\alpha |\mathcal{Y}) \propto \ln p( {\mathcal{Y}} | U,V,R,\alpha )+ \ln p( U|\alpha_{U}) + \ln p( V|\alpha_{V})+ \ln p(R |\mu_{T}, \alpha_{U})+ C
\end{aligned}
\end{equation}
where $C$ is a constant.

With the fixed values of hyperparameters $\{\alpha,\alpha_{U},\alpha_{V},\alpha_{R} \}$, maximizing the posterior conditional distribution over the model parameters $\{U,V,R,\alpha \}$ is equivalent to minimizing the following regularized weighted error function \cite{s:lpp}:
\begin{equation}
\begin{aligned}
\begin{split}
E = \frac{1}{2} \sum\limits_{i=1}^N \sum\limits_{j=1}^N \sum\limits_{t=1}^T \mathcal{I}_{ijt} \Big( \mathcal{Y}_{ijt}-\big( \sum\limits_{d=1}^D U_{id}V_{jd}R_{td}\big) \Big)^2 \\
+ \frac{\gamma_{U}}{2} \sum\limits_{i=1}^N \parallel U_{i} \parallel^2_F + \frac{\gamma_{V}}{2} \sum\limits_{j=1}^N \parallel V_{j} \parallel^2_F +\frac{\gamma_{R}}{2} \sum\limits_{t=1}^T \parallel R_{t} \parallel^2_F
\end{split}
\end{aligned}
\end{equation}
where $\gamma_{U} = \alpha_{U}/ \alpha $, $\gamma_{V} = \alpha_{V}/ \alpha$, $\gamma_{R} = \alpha_{R}/ \alpha$, $\|\cdot\|_F$ denotes the Frobenius norm. For optimizing the latent factors in the error function, we can use the Polak-Ribi\`{e}re variant of non-linear conjugate gradient based method \cite{j:fast} to find a local optimal solution for $\{U,V,R,\alpha \}$. The inference procedure is outlined in Algorithm 1, more details can be referred to \cite{s:lpp}.

\begin{algorithm}
\caption{Conjugate Gradient method for PLTF model }
\label{alg1}
\begin{algorithmic}
\STATE \textbf{Initialize} latent factor parameters $\{\mathbf{U}_{0}, \mathbf{V}_{0}, \mathbf{R}_{0}\}$
\REPEAT
\FOR{$i=1$ to $N$}
\STATE ${U}_{i}^{*} \leftarrow U_{i} + \alpha (\beta \Lambda_{U} - \frac{\partial{\textit{E}}}{\partial U_{i}})$
\ENDFOR
\FOR{$j=1$ to $N$}
\STATE ${V}_{j}^{*} \leftarrow V_{j} + \alpha (\beta \Lambda_{V} - \frac{\partial{\textit{E}}}{\partial V_{j}})$
\ENDFOR
\FOR{$t=1$ to $T$}
\STATE ${R}_{t}^{*} \leftarrow R_{t} + \alpha (\beta \Lambda_{R} - \frac{\partial{\textit{E}}}{\partial R_{t}})$
\ENDFOR
\UNTIL{stopping criterion is met}
\STATE \textbf{Return} $\{\mathbf{U}^{*},\mathbf{V}^{*},\mathbf{R}^{*}\}$
\end{algorithmic}
\end{algorithm}

\subsubsection{Limitations of MAP Estimation}
After obtaining the estimates for latent factors $\{ {U}^{*}, {V}^{*}, {R}^{*}\}$, we may predict the missing link patterns, which means simultaneously inferring multiple relation types between object pairs by Equation (7). However, there are still limitations on the results of MAP estimation. First, since the observed data is sparse and the MAP estimation chooses a single point $\{ {U}^{*}, {V}^{*}, {R}^{*}\}$, there is no model averaging which may lead to higher variance in final prediction, and point estimators also ignore uncertainty in the model parameters $\{{U}, {V}, {R}\}$.

Another limitation with the basic model is about manually tuning the values of the parameters $\{\alpha,\alpha_{U},\alpha_{V},\alpha_{R},\mu_{T}\}$. We can consider a set of appropriate prior values to learn the model parameters, and select the best ones based on cross-validation. However, this approach is infeasible and computationally expensive. Therefore, in the next section we will introduce a fully Bayesian treatment to the proposed PLTF model to avoid its drawback in model parameter learning.

\section{Hierarchical Bayesian Treatment for PLTF Model}
The aforementioned inference procedure for learning latent factors by MAP estimation always leads to overfitting when the model parameters are not properly selected, particularly on large-scale and sparse datasets. Thus we develop an alternative solution which employs the hierarchical Bayesian learning for the PLTF model. Based on the Bayesian treatment, we can make the posterior distribution integrating out all model parameters and hyperparameters, which makes the predictive distribution less likely overfits the observed data and generalizes well on the missing data.

\subsection{Hierarchical Bayesian Model}
As in the PLTF model, the observed data are modeled using a multivariate Gaussian likelihood given by the latent tensor factorization in Equation (2), then the probability of the observed data can be generated from the following generative process:
\begin{equation}
\begin{aligned}
{\mathcal{Y}_{ijt}}| U,V,R, \alpha \sim \mathcal{N} \Big( \big(\sum\limits_{d=1}^D U_{id}V_{jd}R_{td} \big), \alpha^{-1}\Big)
\end{aligned}
\end{equation}

With the consideration of hierarchical Bayesian generative model, we select a conjugate prior on the precision $\alpha$, namely Gamma distribution with shape and scale parameters $l$ and $\theta$ as follows:
\begin{equation}
\begin{aligned}
p(\alpha | l,\theta) = \textbf{Gamma}(l, \theta )= \frac{\theta^{l}}{\Gamma(l)}\alpha_{l-1}exp(\frac{-x}{\theta})
\end{aligned}
\end{equation}

Moreover, we choose the prior distributions for the latent factors of objects and relation types as multivariate Gaussian distributions with mean vector $\mu$ and precision matrix $\Lambda$ (inverse of the covariance matrix):
\begin{equation}
\begin{aligned}
p({U} \mid \mu_{U},\Lambda_{U}) = \prod\limits_{i=1}^N \mathcal{N}(U_{i} \mid \mu_{U},\Lambda^{-1}_{U})
\end{aligned}
\end{equation}

\begin{equation}
\begin{aligned}
p(V \mid \mu_{V},\Lambda_{V}) = \prod\limits_{j=1}^N \mathcal{N}(V_{j} \mid \mu_{V},\Lambda^{-1}_{V})
\end{aligned}
\end{equation}

\begin{equation}
\begin{aligned}
p(R \mid \mu_{R},\Lambda_{R}) = \prod\limits_{t=1}^T \mathcal{N}(R_{t} \mid \mu_{R},\Lambda^{-1}_{R})
\end{aligned}
\end{equation}

Thus, the above prior distributions on the latent factors $\{U,V,R, \alpha \}$ are conjugate to the Gaussian likelihood in the Equation (10). Then we still need to select the prior distributions for the parameters $\{\mu, \Lambda \}$ of the latent factors. Considering to set the multivariate Gaussian parameters and to facilitate subsequent learning procedure, we choose the appropriate conjugate priors as Gaussian-Wishart distributions for the means $\mathbf{\mu}$ and precision matrices $\mathbf{\Lambda}$:
\begin{equation}
\begin{aligned}
\begin{split}
p(\mu_{U},\Lambda_{U})  = p(\mu_{U} | \Lambda_{U}) p(\Lambda_{U})  = \mathcal{N}\big( \mu_{U}|\mu_{0}, ( \kappa_{0} \Lambda_{U} )^{-1} \big) \mathcal{W}(\Lambda_{U}|W_{0},\nu_{0})
\end{split}
\end{aligned}
\end{equation}

\begin{equation}
\begin{aligned}
\begin{split}
p(\mu_{V},\Lambda_{V}) = p(\mu_{V} | \Lambda_{V}) p(\Lambda_{V})  = \mathcal{N}\big( \mu_{V}|\mu_{0}, ( \kappa_{0} \Lambda_{V} )^{-1} \big) \mathcal{W}(\Lambda_{V}|W_{0},\nu_{0})
\end{split}
\end{aligned}
\end{equation}

\begin{equation}
\begin{aligned}
\begin{split}
p(\mu_{R},\Lambda_{R}) = p(\mu_{R} | \Lambda_{R}) p(\Lambda_{R}) = \mathcal{N}\big( \mu_{R}|\mu_{0}, ( \kappa_{T} \Lambda_{R} )^{-1} \big) \mathcal{W}(\Lambda_{R}|W_{0},\nu_{0})
\end{split}
\end{aligned}
\end{equation}
where $\mathcal{W}$ is the Wishart distribution of a $D \times D$ random matrix $\Lambda$ with a scale matrix $W_{0}$ and degrees of freedom $\nu_{0}$.

\begin{figure}[t]
\centering
\includegraphics[width=3in]{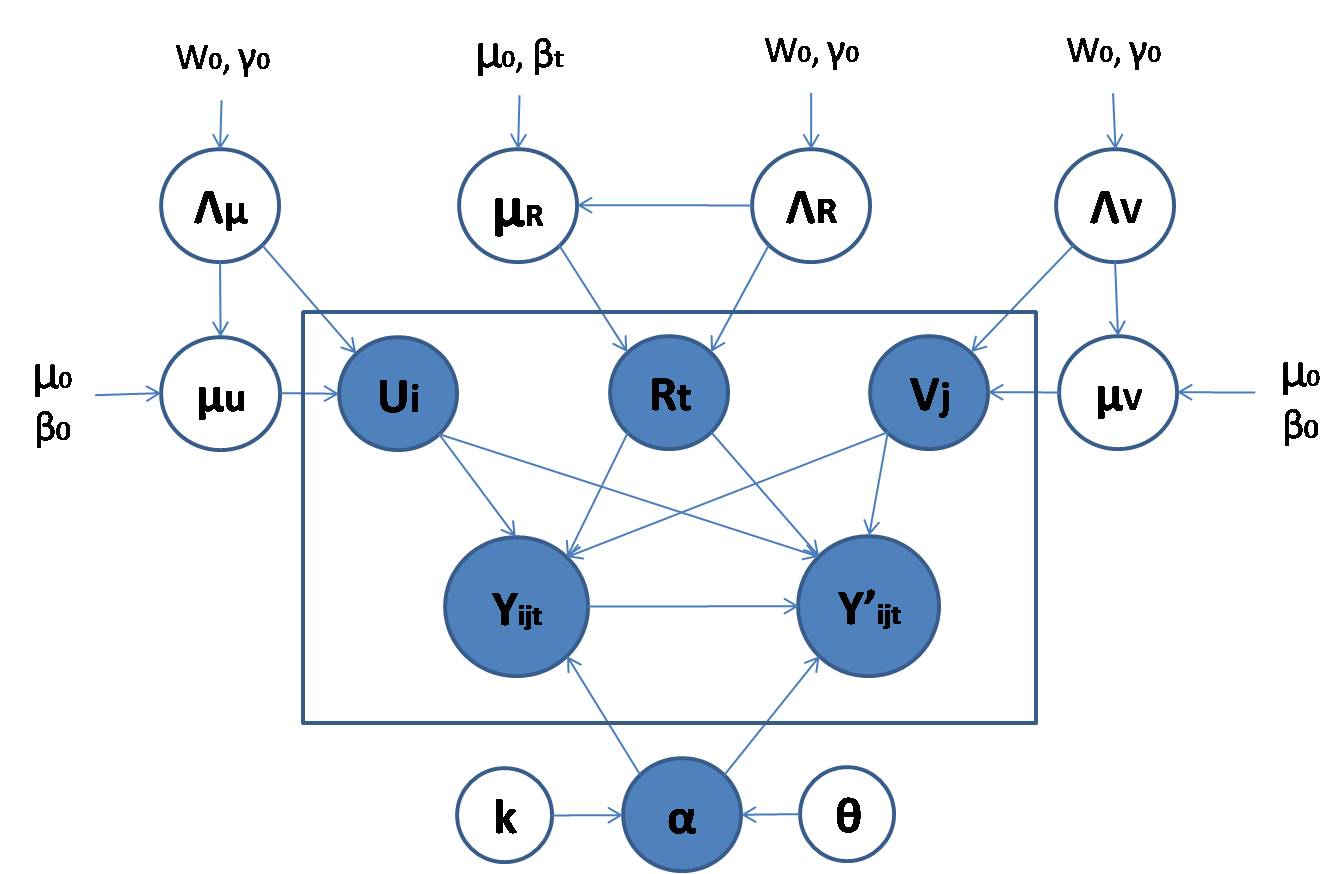}
\caption{Graphical representation for the Hierarchical Bayesian treatment of Probabilistic Latent Tensor Factorization model. Shaded nodes $\{U,V,R,\alpha \}$ indicate the latent factor parameters, $\mathcal{Y}_{ijt}$ and $\mathcal{Y}_{ijt}^{*}$ denote the probability of observed relation and the missing one. Hollow nodes denote the parameters $\{\mu_{U},\mu_{V},\mu_{R}, \Lambda_{U},\Lambda_{V},\Lambda_{R} \}$ and hyperparameters $\{l, \theta,\mu_{0},\kappa_{T},\kappa_{0},W_{0},\nu_{0}\}$. Weight indicator is elided.}
\end{figure}

Figure 3 contains a graphical representation of our proposed Hierarchical Bayesian model. The predictive distribution of the missing link patterns over all the model parameters and hyperparameters conditioned on the observed data can be obtained as follows:
\begin{equation}
\begin{aligned}
\begin{split}
p(\mathbf{\mathcal{Y}^{*}}|\mathbf{\mathcal{Y}}) = \int & p(\mathbf{\mathcal{Y}^{*}}| U,V,R,\alpha) p(U,V,R,\alpha | \mathbf{\mu}, \mathbf{\Lambda}, \mathcal{Y})  p(\mathbf{\mu}, \mathbf{\Lambda}| \mathbf{\Theta}) d \{U,V,R,\alpha,\mathbf{\mu},\mathbf{\Lambda} \}
\end{split}
\end{aligned}
\end{equation}
where $\mathbf{\mu}$ denotes $\{ \mu_{U},\mu_{V},\mu_{R}\}$, $\mathbf{\Lambda}$ denotes $\{\Lambda_{U},\Lambda_{V},\Lambda_{R}\}$, $\mathbf{\Theta}$ denotes the hyperparameters $\{k, \theta,\mu_{0},\kappa_{T},\kappa_{0},W_{0},\nu_{0}\}$ for the conjugate prior distributions which are specified in the Bayesian learning.

Based on the hierarchical Bayesian treatment, the modified PLTF model can marginalize over all model parameters and hyperparameters, and make more efficient performance. Moreover, Bayesian learning can tune the values of hyperparameters $\mathbf{\Theta}$ in a reasonably set of model space, which has little change on the stable performance of our model. We refer to the resulting model as Hierarchical Bayesian Probabilistic Latent Tensor Factorization (HB-PLTF) model, and we will show an efficient inference procedure for estimating model parameters and hyperparameters in the next subsection.

\subsection{Inference with Markov Chain Monte Carlo}
The exact solution to the predictive prediction in Equation (18) is analytically intractable due to the difficulty of computing the posterior distribution. We thus resort to the approximate inference. In this paper, we employ the sampling-based method, i.e., Markov Chain Monte Carlo (MCMC) \cite{r:pmf} \cite{m:nft} technique.

From the MCMC sampling-based inference view, we can draw the approximation to the predictive distribution of Equation (18) as follows:
\begin{equation}
\begin{aligned}
p( {\mathcal{Y}^{*}}|{\mathcal{Y}}) \approx \frac{1}{K} \sum \limits_{k=1}^K p({\mathcal{Y}^{*}}|{U}^{(k)}, {V}^{(k)}, {R}^{(k)}, \alpha^{(k)})
\end{aligned}
\end{equation}
where the prediction can be approximated by the expectation of $p({\mathcal{Y}^{*}} |{U}^{(k)}, {V}^{(k)}, {R}^{(k)},\alpha^{(k)})$ by a sequence of samples $\{ {U}^{(k)}, {V}^{(k)}, {R}^{(k)},\alpha^{(k)} \}$ drawn from a Markov chain whose proposal distribution is $p(U,V,R,\alpha |\mathcal{Y})$ that denotes the posterior distribution over the model parameters.

In our work, we use Gibbs sampling, one of the most widely used MCMC techniques, by which we can draw the conditional distributions on latent variables that have a parametric form easily to be sampled from \cite{c:mc}. During the procedure of Gibbs sampling, the latent variables of model parameters and hyperparameters are partitioned into several blocks, and each block of latent variables is sampled iteratively given some initial value of the parameters while all the others are fixed until convergence.

To apply the Gibbs sampling method in our Bayesian probabilistic model we need to derive the appropriate conditional distributions based on the predictive distribution of Equation (18). The use of conjugate priors for the model parameters and hyperparameters makes the posterior conditional distributions easy to sample from, which leads to an efficient sampling procedure. Here according to the Bayesian rules and Equation (18) we can derive the posterior distribution conditioned on the observed data as follows:
\begin{equation}
\begin{aligned}
p (U,V,R,\alpha | \mathcal{Y}) \propto p(\mathcal{Y}| U,V,R,\alpha) p(\alpha |k,\theta) p(U|\mu_{U},\Lambda_{U})p(V|\mu_{V},\Lambda_{V})\\
p(R| \mu_{R},\Lambda_{R}) p(\mu_{U},\Lambda_{U}) p(\mu_{V},\Lambda_{V}) p(\mu_{R},\Lambda_{R})
\end{aligned}
\end{equation}
The whole Gibbs sampling procedure is illustrated in Algorithm 2.

\subsubsection{Posterior Distributions of Latent Factors}
To learning the latent feature factor $U^{*}$, we can derive the posterior conditional distribution of the corresponding samples which follows the multivariate Gaussian distribution conditioned on the other latent factors and the observed data as follows:
\begin{equation}
\begin{aligned}
p(U^{*} |V^{*},R^{*},\mathcal{Y}, \mu_{U},\Lambda_{U}) = \prod\limits_{i=1}^N \mathcal{N}(U_{i}^{*} | \mu^{*}_{i},(\Lambda^{*}_{i})^{-1})
\end{aligned}
\end{equation}
with the posterior precision matrix and the mean as:
\begin{equation*}
\begin{aligned}
\Lambda^{*}_{i} = \Lambda_{U} + \alpha \sum \limits_{j=1}^N \sum \limits_{t=1}^T \mathcal{I}_{ijt} (V_{j}^{*} \bullet R_{t}^{*})^{T}((V_{j}^{*} \bullet R_{t}^{*}))
\end{aligned}
\end{equation*}
\begin{equation*}
\begin{aligned}
\mu^{*}_{i} = (\Lambda^{*}_{i})^{-1} \big( \mu_{U} \Lambda_{U} + \alpha \sum \limits_{j=1}^N \sum \limits_{t=1}^T \mathcal{I}_{ijt} (V_{j}^{*} \bullet R_{t}^{*}) \mathcal{Y}_{ijt} \big)
\end{aligned}
\end{equation*}
where the symbol $" \bullet "$ denotes the element-wise product of two vectors. Similarly the posterior conditional distribution for the latent factor $V^{*}$ can be drawn in the same manner and have the similar form as for $U^{*}$.
\begin{algorithm}[t]
\caption{Gibbs sampling for Bayesian PLTF model}
\label{alg2}
\begin{algorithmic}
\STATE \textbf{Initialize} latent factor parameters $\{\mathbf{U}_{0}, \mathbf{V}_{0}, \mathbf{R}_{0}\},\alpha_{0}$
\FOR{$k=1$ to $K$}
\STATE Sample the model parameter and hyperparameters based on Equation (23),(24),(25), respectively:
\STATE $ \alpha^{(k)^{*}} \leftarrow \text{sample from} \quad Gamma(l^{*}, \theta^{*}|U^{(k)},V^{(k)},R^{(k)},\mathcal{Y})$

\STATE $\{\mu_{U}^{(k)},\Lambda_{U}^{(k)}\}  \leftarrow \text{sample from} \quad p(\mu_{U}^{(k)},\Lambda_{U}^{(k)}|U^{(k)}) $

\STATE $\{\mu_{V}^{(k)},\Lambda_{V}^{(k)}\} \leftarrow \text{sample from} \quad p(\mu_{V}^{(k)},\Lambda_{V}^{(k)}|V^{(k)}) $

\STATE $\{\mu_{R}^{(k)},\Lambda_{R}^{(k)}\} \leftarrow \text{sample from} \quad p(\mu_{R}^{(k)},\Lambda_{R}^{(k)}|R^{(k)})$

\FOR{$i=1$ to $N$}
\STATE $U_{i}^{(k+1)^{*}} \leftarrow \quad p(U^{*}_{i} |V^{(k)^{*}},R^{(k)^{*}},\alpha^{(k)^{*}},\mathcal{Y},\mu_{U}^{(k)},\Lambda_{U}^{(k)})$
\ENDFOR

\FOR{$j=1$ to $N$}
\STATE $V_{j}^{(k+1)^{*}} \leftarrow \quad p(V^{*}_{j} |U^{(k+1)^{*}},R^{(k)^{*}},\alpha^{(k)^{*}},\mathcal{Y},\mu_{V}^{(k)},\Lambda_{V}^{(k)})$
\ENDFOR

\FOR{$t=1$ to $T$}
\STATE $R_{t}^{(k+1)^{*}} \leftarrow \quad p(R^{*}_{t} |U^{(k+1)^{*}},V^{(k+1)^{*}},\alpha^{(k)^{*}},\mathcal{Y},\mu_{R}^{(k)},\Lambda_{R}^{(k)})$
\ENDFOR

\ENDFOR
\STATE \textbf{Return} $\{\mathbf{U}^{*},\mathbf{V}^{*},\mathbf{R}^{*},\alpha^{*}, \mathbf{\mu}, \mathbf{\Lambda} \}$
\end{algorithmic}
\end{algorithm}

As for the latent factor $R^{*}$ of relation types, we assume it is influenced by the interaction between two latent feature factors $U^{*}$ and $V^{*}$ with respect to the object features, and then we still consider its posterior conditional distribution with the same parametric form as its prior distribution which follows the multivariate Gaussian distribution:
\begin{equation}
\begin{aligned}
p(R^{*} |U^{*},V^{*},\mathcal{Y}, \mu_{R},\Lambda_{R}) = \prod\limits_{t=1}^T \mathcal{N}(R_{t}^{*} | \mu^{*}_{t},(\Lambda^{*}_{t})^{-1})
\end{aligned}
\end{equation}
where
\begin{equation*}
\begin{aligned}
\Lambda^{*}_{t} = \Lambda_{R} + \alpha \sum \limits_{i=1}^N \sum \limits_{j=1}^N \mathcal{I}_{ijt} (U_{i}^{*T}U_{i}^{*}) \bullet (V_{j}^{*T}V_{j}^{*})
\end{aligned}
\end{equation*}
\begin{equation*}
\begin{aligned}
\mu^{*}_{t} = (\Lambda^{*}_{t})^{-1} \big( \mu_{R} \Lambda_{R} + \alpha \sum \limits_{i=1}^N \sum \limits_{j=1}^N \mathcal{I}_{ijt} (U_{i}^{*} \bullet V_{j}^{*}) \mathcal{Y}_{ijt} \big)
\end{aligned}
\end{equation*}

\subsubsection{Posterior Distributions of Model Parameters}
From the posterior distribution form with the aforementioned conjugate priors, we can derive the posterior conditional distribution for samples $\{\alpha^{*}\}$ which follow the Gamma distribution as:
\begin{equation}
\begin{aligned}
p(\alpha^{*} | l^{*},\theta^{*} ) = \textbf{Gamma}(l^{*}, \theta^{*})
\end{aligned}
\end{equation}
where
\begin{equation*}
\begin{aligned}
l^{*} = l + \frac{1}{2} \sum\limits_{i=1}^N \sum\limits_{j=1}^N \sum\limits_{t=1}^T \mathcal{I}_{ijt}
\end{aligned}
\end{equation*}
\begin{equation*}
\begin{aligned}
\theta^{*} = \Big ( \theta^{-1} + \frac{1}{2} \sum\limits_{i=1}^N \sum\limits_{j=1}^N \sum\limits_{t=1}^T \mathcal{I}_{ijt} \big(\mathcal{Y}_{ijt}-(\sum\limits_{d=1}^D U_{id}V_{jd}R_{td})\big)^2 \Big)^{-1}
\end{aligned}
\end{equation*}

Note that for obtaining the conditional distributions over the latent factors, we need to derive the posterior conditional sampling for the model hyperparameters, i.e., the means $\{\mathbf{\mu }\}$ and the precision matrices $\{ \mathbf{\Lambda } \}$ simultaneously.

Considering the conjugate prior distribution of $\{ \mu_{U},\Lambda_{U} \}$ in Equation (15), we have the conditional distribution with the form of Gaussian-Wishart distribution as follows:
\begin{equation}
\begin{aligned}
p(\mu_{U},\Lambda_{U}|\mathbf{U}) = \mathcal{N}\big( \mu_{U}|\mu_{0}^{*}, ( \kappa_{0}^{*} \Lambda_{U})^{-1} \big) \mathcal{W}(\Lambda_{U}|W_{0}^{*},\nu_{0}^{*})
\end{aligned}
\end{equation}
with
\begin{equation*}
\begin{aligned}
\mu_{0}^{*}= \frac{\kappa_{0}\mu_{0}+ N \bar{U}}{\kappa_{0}+ N} , \kappa_{0}^{*}=\kappa_{0} + N, \nu_{0}^{*}= \nu_{0}+ N;
\end{aligned}
\end{equation*}
\begin{equation*}
\begin{aligned}
(W_{0}^{*})^{-1}= W_{0}^{-1} + C + \frac{\kappa_{0}N}{\kappa_{0}+ N}(\bar{U}-\mu_{0})(\bar{U}-\mu_{0})^{T}
\end{aligned}
\end{equation*}
\begin{equation*}
\begin{aligned}
C= \sum_{i=1}^N (U_{i}-\bar{U})(U_{i}-\bar{U})^{T}
\end{aligned}
\end{equation*}
where $\bar{U}=\frac{1}{N}\sum_{i=1}^N U_{i}$ is the sample mean. Similarly, the conditional distribution for $\{ \mu_{V},\Lambda_{V} \}$ has the same form.

As for the hyperparamters $\{ \mu_{R},\Lambda_{R} \}$ for the latent feature factor of relation types, we can still derive the similar parametric form of Gaussian-Wishart distribution as follows:
\begin{equation}
\begin{aligned}
p(\mu_{R},\Lambda_{R}|\mathbf{R}) = \mathcal{N}\big( \mu_{R}|\mu_{0}^{*}, ( \kappa_{T}^{*} \Lambda_{R})^{-1} \big) \mathcal{W}(\Lambda_{R}|W_{0}^{*},\nu_{0}^{*})
\end{aligned}
\end{equation}
with a little difference on $\kappa_{T}^{*}=\kappa_{T} + T$, the mean $\mu_{0}^{*}$ and the precision matrix $W_{0}^{*}$ have the forms respectively as:
\begin{equation*}
\begin{aligned}
\mu_{0}^{*}= \frac{\kappa_{T}\mu_{0}+ T \bar{R}}{\kappa_{T}+ T}, \nu_{0}^{*}= \nu_{0}+ T, S= \sum_{t=1}^T (R_{t}-\bar{R})(R_{t}-\bar{R})^{T};
\end{aligned}
\end{equation*}
\begin{equation*}
\begin{aligned}
(W_{0}^{*})^{-1}= W_{0}^{-1} + S + \frac{\kappa_{T}T}{\kappa_{T}+ T}(\bar{R}-\mu_{0})(\bar{R}-\mu_{0})^{T}
\end{aligned}
\end{equation*}
where $\bar{R}=\frac{1}{T}\sum_{t=1}^T R_{t}$ denotes the sample mean.

\subsection{Computational Complexity Analysis}
In this subsection, we discuss the computational complexity of our proposed algorithms in the implementation. For per iteration the proposed non-Bayesian PLTF model and the Hierarchical Bayesian model both require $O(2ND^{3}+ TD^{3}+ |\mathcal{Y}|D^{2})$ computational time, where $|\mathcal{Y}|$ denotes the number of observed relations in the learning phrase. For the choice of $D$, we can choose the values according to the tradeoff between the model complexity and the learning efficiency. For the hyperparameters, the proposed Hierarchical Bayesian model can eliminate the complexity of manual adjustment by introducing the prior distributions for them. For the Gibbs sampling procedure, since the convergence of sampling usually takes a long time, the MAP results from PLTF model can be used to initialize the Gibbs sampling. Moreover, we can stop sampling when the accuracy archives the desirable level in the experiments.

\section{Experiments}
We evaluate the performance of our proposed models on discovering missing link patterns on three real-world multi-relational datasets and make some discussions about the results later.

\subsection{Datasets}
In the experiments, we examine how our proposed models behave on real-world multi-relational networks. Three datasets are collected: Kinship relational dataset, Country relational dataset and YouTube social dataset.
\begin{itemize}
\item \textbf{Kinship relational dataset.} The Kinship relational dataset consists of kinship relationships among the members of the ALyawarra tribe \cite{den:kinship}. We extract 26 relation types among 104 people in the dataset (such as "father" or "wife" relations), and then build the multi-relational network. For evaluation, we construct the tensor model of size ${\textit{104}}{\times}{\textit{104}}{\times}{\textit{26}}$.

\item \textbf{Country relational dataset.} The Country relational dataset consists of international relations among different countries in the world \cite{rum:nations}. We extract 56 relation types among 14 countries in the dataset (such as "emigrants" or "exports"), and then build the multi-relational network for this dataset. For evaluation, we construct the tensor model of size ${\textit{14}}{\times}{\textit{14}}{\times}{\textit{56}}$.

\item \textbf{Youtube social dataset.} YouTube is currently the most popular video sharing web site, which allows users to interact with each other in multiple relations such as contacts, subscriptions or sharing favorite videos \footnote{$http://www.public.asu.edu/~ltang9$}. we choose 3,000 active user profiles in the network, and construct the tensor model of size ${\textit{3000}}{\times}{\textit{3000}}{\times}{\textit{5}}$ with five types of relations.
\end{itemize}

\subsection{Experimental Setup}
We apply our proposed PLTF model and the hierarchical Bayesian version for Link Pattern Prediction (LPP) problem. In addition, we test the Bayesian treatment of probabilistic latent factorization model by initializing the Gibbs sampler with either random latent factor matrices (denoted as "HB-rPLTF") or the latent factors obtained by training the PLTF model (denoted as "HB-tPLTF"). We also compare these methods with the other state-of-the-art methods.
\begin{itemize}
\item LPP-LFRM model (Latent Feature Relational Model) \cite{kurt:lfrm}. This is a nonparametric latent feature relational model which infers a global set of latent binary features for each object as well as how those latent features interact in the multi-relational networks.

\item LPP-BPMF model (Bayesian Probabilistic Matrix Factorization) \cite{r:pmf}. This is a probabilistic matrix factorization model applied separately to each relation slice $\mathcal{Y}_{::t}$ of the original $\mathcal{Y}$ tensor data. Each relation within this procedure is thus handled independently of the other relations in the network. We implement and report the performance of this model in the LPP problem. Comparing our PLTF model with this mono-relational method will allow to examine the benefit, if any, of multi-relational prediction.
\end{itemize}

Each tube fiber in our PLTF tensor model can represent a link pattern between two objects. For the experiments, we choose a given fraction (e.g., $20\%$) of the tube fibers as unobserved link patterns for the test data. For the Hierarchical Bayesian treatment of our proposed PLTF model, parameters are set according to prior knowledge without tuning: $\mu_{0} = 0$, $\nu_{0}=D$, $W_{0}= \mathbf{I}$, $k=5$, $\theta=1$, $\kappa_{0}=2$, $\kappa_{T}=1$. The algorithms are implemented in Matlab. Then, we use the AUC (score Area Under the receiver operating characteristic Curve), which is a robust measure for sparse dataset \cite{Acar:Link}, as the evaluation metric to test the link pattern prediction performance. For the experiments we evaluate the methods by repeating the process five times and report the average results.
\begin{table}[t]
\centering
\begin{tabular*}{\linewidth}{@{\extracolsep{\fill}}lcc}
\hline\hline
         & Kinship Dataset & Countries Dataset  \\ \hline
HB-tPLTF & 0.9483          & 0.9187             \\
HB-rPLTF & 0.9401          & 0.9111             \\
PLTF     & 0.9269          & 0.8994             \\
LPP-LFRM & 0.9183          & 0.8772             \\
LPP-BPMF & 0.8022          & 0.7827             \\ \hline
  \end{tabular*}
\caption{Average AUC Performances with different methods on Kinship and Countries datasets.}
\end{table}

\subsection{Experimental Results}
We first compare our proposed PLTF model and its Bayesian variants to the state-of-the-art model LFRM and the mono-relational model LPP-BPMF. Table 1 reports the average AUC performances of these models on the Kinship and Countries datasets. For the PLTF model, we can learn the model by maximizing the posterior distribution, and set the regularization terms $\gamma_{U}=\gamma_{V}=\gamma_{R}= 0.01$. For the HB-rPLTF model and HB-tPLTF model we generate 300 samples in sampling when the results stabilizes as well as in LPP-BPMF model.

From Table 1, the best performing method among all the models are the HB-PLTF variants. And the multi-relational models (HB-tPLTF, HB-rPLTF, PLTF and LPP-LFRM) consistently have better results than the LPP-BPMF mono-relational method in our prediction task on the two multi-relational datasets. This seems to show that multi-relational prediction methods allow capturing the correlations among multiple relation types so as to improve the accuracy of link pattern prediction. In contrast, the LPP-BPMF based method only processes the target relation type separately without considering the additional cross link pattern information.

Another observation is that our proposed PLTF model and its Bayesian variants perform better than the LFRM model. The reason is that the proposed PLTF models can more efficiently explore the impact of different relations by introducing the latent factor for multiple relation types and capturing the interactions between three latent factors of objects and relation types, while the LFRM model just infers a global set of binary latent feature matrices for all relations and only considers the latent feature factor of objects.

Table 1 also shows the performance comparison between PLTF model and its Hierarchical Bayesian versions. We can observe that HB-PLTF versions outperform the non-Bayesian PLTF model, which indicates the prediction performance can be enhanced by integrating out model parameters and hyperparameters and by the efficient procedure of Gibbs sampling in the model space. However, the difference between the Hierarchical Bayesian versions with the random initialization and with the PLTF initialization is inconspicuous which shows the stability of the Bayesian version of PLTF model.

Next, we compare the prediction quality of our proposed PLTF models to the LPP-BPMF mono-relational model on YouTube dataset, which is a large-scale and sparse dataset. Results about average AUC performance for those models with varying percentages (e.g. $20\%, 40\%$, and $60\%$) of missing link patterns are indicated in Table 2. As we can see, HB-tPLTF provides the best prediction quality among all the methods. Moreover, the variations of three PLTF models clearly outperforms the LPP-BPMF mono-relational model, the results also confirm the ability of our PLTF models to deal with the multi-relational data and to capture the correlations among multiple relations, which can be used to improve the accuracy of link pattern prediction.

\begin{table}[t]
\centering
\begin{tabular*}{\linewidth}{@{\extracolsep{\fill}}lccc}
\hline\hline
         & $20\%$          & $40\%$           & $60\%$   \\ \hline
HB-tPLTF & 0.9101          & 0.8348           & 0.8107   \\
HB-rPLTF & 0.9067          & 0.8210           & 0.8001   \\
PLTF     & 0.8740          & 0.7999           & 0.7512   \\
LPP-BPMF & 0.7202          & 0.6575           & 0.6101   \\ \hline
\end{tabular*}
\caption{Average AUC Performances with varying percentages of missing link patterns on YouTube dataset}
\end{table}

\begin{figure}[t]
\centering
\includegraphics[width=2.8in]{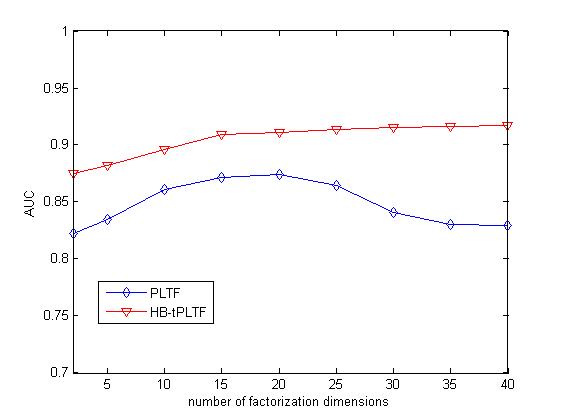}
\caption{The impact of latent tensor factorization dimensions on YouTube dataset}
\end{figure}
\subsubsection{Impact of Latent Factorization Dimensions}
Parameter \textit{D} implies the number of latent tensor factorization dimensions which also determines both the number of model parameters and model complexity. We conduct experiments on the three datasets to examine how the predictive performance is affected by \textit{D}.

Figure 4 shows the impact of the latent factorization dimensions on the performance of our proposed PLTF model and its Hierarchical Bayesian variations on the YouTube dataset. Frow Figure 4 we can observe that with the increasing number of factorization dimensions the prediction performance of the PLTF model does not improve and sometimes even becomes overfitting, while the performance of HB-tPLTF model nearly remain stable. The results again indicate the advantage of Bayesian treatment for the probabilistic model. Considering the tradeoff between the model complexity and the learning efficiency We thus select the optimal number of factorization dimension for YouTube dataset is $D=20$.

We omit the detailed results for the other datasets as the similar trend of choosing model dimensions can be observed. For the Kinship dataset we can infer the optimal number factorization dimensions is $D=11$, and for the Countries dataset $D=7$.

\subsubsection{Impact of Various Relation Types}
In multi-relational networks objects connect with each other which follows some kind of interaction pattern within the context of certain specific relation type. Different relation types demonstrate distinct interaction patterns in the network, which is captured by the \textit{link pattern} containing multiple relation types. Here we try to exploit the effect of various relation types on the prediction performance, taking YouTube dataset as example.

YouTube dataset contains five relation types among which \textit{Contact relation} is the most sparse one while the other relational contexts are denser \cite{Tang:Uncoverning}. We consider the setting of incorporating some specific relation type used in the training set when the $20\%$ fraction is selected as test data. The higher performance the result demonstrate, the more influence the specific relation type has on the link pattern prediction.
\begin{figure}[t]
\centering
\includegraphics[width=3in]{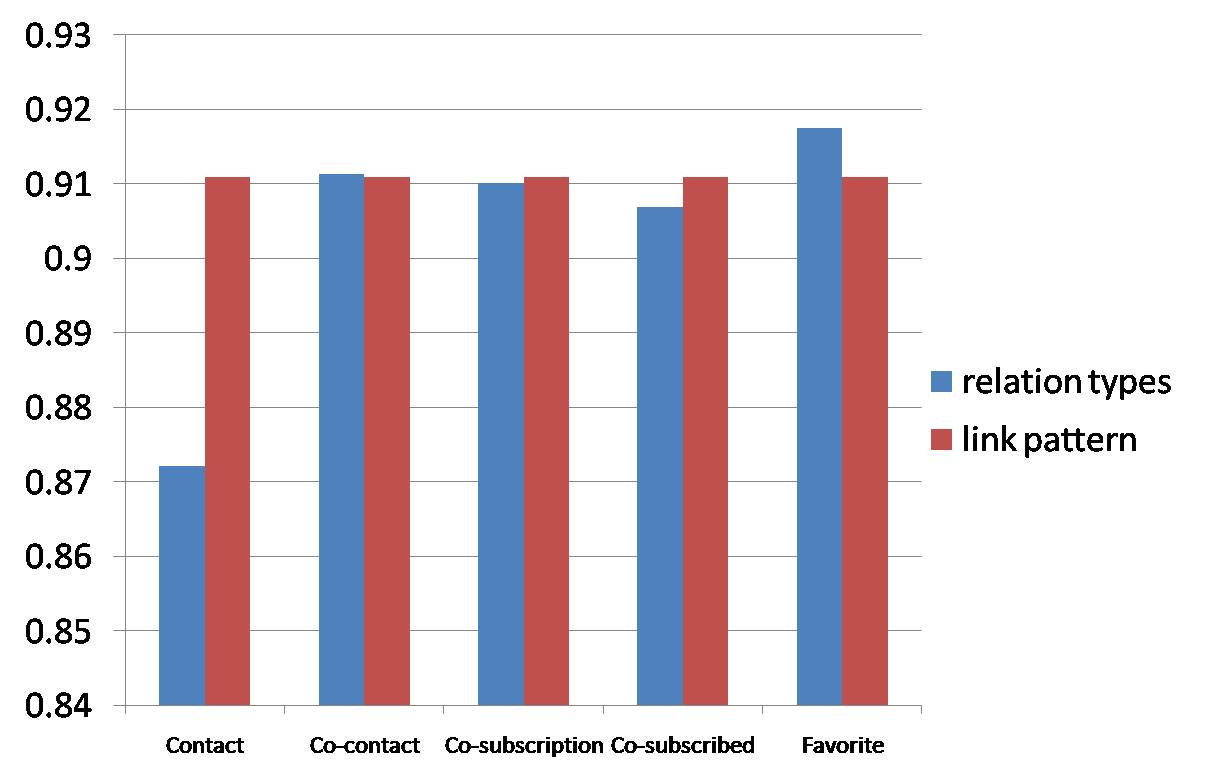}
\caption{Impact of various relation types on the prediction performance of PLTF model by incorporating each relation used in the training set compared with the normal result on YouTube dataset}
\end{figure}
Figure 5 shows the average prediction results. We can observe that the \textit{Favorite relation} help boost the performance significantly while the \textit{Contact relation} has the least impact on the results, which is consistent with our intuition. The results indicate our proposed model can not only capture the correlations among relation types but also reveal the influence of each relation type on the link pattern prediction performance. For example, people who have shared their favorite videos on YouTube are more likely to construct \textit{Contact relation} and subscribe to each other while not vice versa. We can also rank the multiple relation types according to their impact on the prediction performance as follows: $ \textit{Favorite relation} \quad \succ \quad \textit{Co-contact relation} \quad \succ \quad \textit{Co-subscription relation} \quad \succ \quad \textit{Co-subscribed relation} \quad \succ \quad \textit{Contact relation}$.

\section{Conclusions and Future Work}
In this paper, we have proposed a new task of Link Pattern Prediction (LPP) problem and then developed a Probabilistic Latent Tensor Factorization (PLTF) model which represents social interaction patterns in multi-relational networks. For constructing the model, we introduce the specific latent factor for different relation types in addition to using latent factors to characterize object features. We also provide the Hierarchical Bayesian treatment of the probabilistic model to avoid overfitting for solving the LPP problem. For that, we derive an efficient Gibbs sampling method to learn the model parameters and hyperparameters. The experiments are conducted on several real world datasets and demonstrate significant improvements over several existing state-of-the-art methods and the ability to capture the correlations among different relation types, reveal the impact of distinct relation types in the multi-relational networks.

There are several directions for future work that we will consider as the extensions of our proposed model. First, it would be interesting to investigate the evolutionary aspect of link patterns in the multi-relational networks over time. Second, we will consider some applications which can use link patterns to improve our understanding of social interaction and large-scale patterns of human association.

\section{Acknowledgments}
We would like to thank Kurt T.Miller for providing the Kinship and Countries datasets.

\bibliographystyle{unsrt}
\bibliography{llncs}

\end{document}